\documentclass[fleqn]{elsart}
\usepackage{epsf,bezier,eepic}

\def\aa{\alpha}
\def\bb{\beta}
\def\cc{\gamma}
\def\<{\langle}
\def\>{\rangle}
\def\be{\begin{equation}}
\def\ee{\end{equation}}

\newcommand{\bm}[1]{\mbox{\boldmath $ #1 $}}
\begin{document}
\begin{frontmatter}

\title{
  Exact Treatment of the Pauli Exclusion Operator
 in Nuclear Matter Calculation
       }

\author[KIT]{K. Suzuki}
\author[KIT]{R. Okamoto}
\author[KDT]{M. Kohno}
\author[Kyoto]{S. Nagata}$^{,}$\footnote{Present address: 20-4 
Miyanowakicho, Syugakuin, Sakyo-ku, Kyoto 606-8061, Japan.}

\address[KIT]{Department of Physics, Kyushu Institute of Technology,
Kitakyushu 804-8550, Japan}
\address[KDT]{Physics Division, Kyushu Dental College, Kitakyushu 803-8580, Japan}
\address[Kyoto]{Department of Applied Physics, Faculty of Engineering, 
Miyazaki University, Miyazaki 889-2192, Japan}

\begin{abstract}
\indent
Exact expressions of the Pauli exclusion operator $Q$ 
in the nuclear matter calculation are presented in detail. 
Exact formulae are also given for the calculations of the 
single-particle-potential energy and the binding energy per nucleon with
the exact $Q$ operator. Numerical calculations of the $G$ matrix
in the lowest-order Brueckner theory are carried out to check
the reliability of the standard angle-average approximation for the
 $Q$ operator by employing the Bonn B and C {\it NN} potentials.
It is observed that the exact treatment of the operator $Q$ brings about
non-negligible and attractive contributions to the binding energy.

\end{abstract}
\end{frontmatter}

\section{Introduction}

\setcounter{equation}{0}

It is one of the fundamental and open problems 
in nuclear structure physics to understand 
the saturation property of nuclear matter starting from a nucleon-nucleon
({\it NN}) interaction\cite{rf:Day67,rf:Spr72,rf:HKO95}.
Because of the presence of strongly repulsive components in the short-range
part of the {\it NN} interaction, the nucleon-nucleon scattering correlation
has a predominant importance in the nuclear matter calculation. 
The Bethe-Goldstone equation has been used for solving
a two-nucleon scattering problem in nuclear matter. 
The influence of  nuclear medium on the nucleon-nucleon scattering
 is taken into account by considering  the Pauli exclusion principle
 and the self-energy effect on scattering nucleons. Therefore the 
accurate treatment of the Pauli exclusion operator has been one of
the essential requirements for the numerical calculation of nuclear matter.

 The Pauli exclusion operator depends, in principle, not only on the
 magnitudes of the total and relative momenta of  
scattering two nucleons but also on their angles. This angular dependence 
leads to  couplings among partial waves, which makes numerical computations 
 difficult. The difficulty due to the angular dependence has been avoided
by employing the angle-average approximation\cite{rf:BG58,rf:HT70}.
The angle-average approximation, however, does not necessarily have 
 quantitative justification, although there have been some studies 
to assess the reliability of this 
approximation\cite{rf:LM63,rf:KS68,rf:IKF80,rf:Leg78,rf:CR89,rf:Sar96}.

One of the purposes of this work is to give  analytic formulae
for the calculation of the nuclear matter binding energy in the exact 
treatment of the Pauli exclusion operator $Q$. Furthermore we want
 to clarify how the $G$ matrix depends on  angular and linear 
momenta. With  analytic expressions of the operator $Q$ and the $G$
 matrix we perform  numerical calculations to investigate the effect 
of the exact treatment of the operator $Q$ on the 
nuclear matter binding energy
by solving the Bethe-Goldstone equation.

Very recently Schiller, M\"uther and Czerski\cite{rf:SMC99} reported their 
calculations of nuclear matter properties using the exact Pauli operator $Q$.
 They pointed out that the standard angle-average approximation for the
operator $Q$ tends to underestimate the binding energy per nucleon at
low densities but overestimate it at higher densities. The exact treatment 
of the operator $Q$ thus leads to a non-negligible improvement in the calculated
saturation points.
 It would be interesting and deserving to examine whether
the similar result can be reproduced in an alternative approach
to the nuclear matter calculation with the exact $Q$ operator. 

The organization of this paper is as follows.
In Section 2 we give an exact and analytic expression of the
operator $Q$ in angular-momentum-coupling states. Section 3 is devoted to derive
 formulae which are necessary for the calculation of the 
single-particle-potential energy and the binding energy per nucleon in the
exact treatment of the operator $Q$. We show, in Section 4, 
 numerical results for the calculations with the exact and
angle-averaged $Q$ operators by employing 
the  Bonn B and C {\it NN} potentials\cite{rf:Bonn}.
We discuss the effect of the exact treatment of the operator $Q$ and 
the accuracy of the angle-average approximation at various 
nuclear matter densities.
 Conclusions obtained in this study are given in Section 5.

\clearpage
\section{
Exact expression of the Pauli exclusion operator
        }
\setcounter{equation}{0}

The reaction matrix ($G$ matrix)  in nuclear matter is defined by the
Bethe-Goldstone equation
 \begin{equation}
  \label{eq:bgeq}
   G= v + v\frac{Q}{e}G,
 \end{equation}
where $v$ is the two-body {\it NN} interaction, $e$ gives the starting energy
 minus the energy of an intermediate two-particle state, and $Q$ 
stands for the Pauli exclusion 
operator. The  operator $Q$ prevents two particles from
scattering into intermediate states
with momenta below the Fermi momentum $k_{\rm F}$, which  is  written as
 \be
  \label{eq:qop1}
   Q=\frac{1}{2}\sum_{\aa\bb}|\aa\bb \> \< \aa\bb |
     \Theta (k_\aa-k_{\rm F})  \Theta (k_\bb-k_{\rm F}),    
 \ee
where $\aa$ is a single-particle state with the  momentum 
{\boldmath $k$}$_{\aa}$,
 the spin projection $s_{\aa}$ and the isospin projection $\tau_{\aa}$.
The state $|\aa\bb\>$ is a normalized and antisymmetrized 
two-nucleon state, and  $\Theta (x) $ the Heaviside step function.
In the relative and center-of-mass (c.m.) momentum system the operator $Q$ is
 written as
\begin{eqnarray}
 \label{eq:qop2}
  Q=\frac{1}{2}\sum_{s_{\aa}\tau_{\aa}s_{\bb}\tau_{\bb}}
     & &\int K^2 dK d{\hat {\bm{K}}}
      \int k^2 dkd{\hat {\bm{k}}}\nonumber\\
     & &\times|\bm{K}\bm{k} s_{\aa}\tau_{\aa}s_{\bb}\tau_{\bb}\> 
       \{\< \bm{K}\bm{k} s_{\aa}\tau_{\aa}s_{\bb}\tau_{\bb} |
       -\< \bm{K}\hspace{1mm}-\bm{k} s_{\bb}\tau_{\bb}s_{\aa}\tau_{\aa} |\}
       \nonumber\\
     & &\times \hspace{1mm}\Theta(|\frac{\bm{K}}{2}+\bm{k}|-k_{\rm F})
               \hspace{1mm}\Theta (|\frac{\bm{K}}{2}-\bm{k}|-k_{\rm F}),
\end{eqnarray}
where {\boldmath $K$}and {\boldmath $k$} are the c.m. and 
relative angular momenta given  by 
{\boldmath $K$}={\boldmath $k$}$_{\aa}+${\boldmath $k$}$_{\bb}$ and 
{\boldmath $k$}=({\boldmath $k$}$_\aa-${\boldmath $k$}$_{\bb})/2$, respectively.

We consider a partial wave decomposition of the relative state 
coupled to the angular momentum {\boldmath $J$} with the c.m. momentum
{\boldmath $K$} as
\begin{eqnarray}
 \label{eq:state}
  |\bm{K}k(lS)J M TT_z\>
   = &f_{l S T}&\sum_{m, S_z} \< l m SS_z | J M\>\nonumber\\
     &\times&\int d{\hat {\bm{k}}} \hspace{1mm}Y_{l m}({\hat {\bm{k}}})
        |\bm{K}\bm{k}\> |SS_zTT_z\>,
\end{eqnarray}
where {\boldmath $S$} is the total spin and {\boldmath $l$} the orbital
 angular momentum  of
two-body relative motion.
We here have introduced  
the anti-symmetrization factor defined by 
$ f_{l S T}=[1-(-1)^{\scriptsize{l+S+T}}]/2$.
The matrix element
of the operator $Q$ between angular-momentum-coupling states becomes 
\begin{eqnarray}
 \label{eq:QMAT}
  \lefteqn{
   \< \bm{K}k(l_1 S_1)J_1 M_{1} T_1 T_{z1}|
           Q
  |\bm{K}'k'(l_2 S_2)J_2 M_{2} T_2 T_{z2}\>
           }\hspace{1cm}\nonumber\\
  &=& \delta (\bm{K}-\bm{K}')\frac{\delta (k-k')}{k^2}
       \delta_{S_1 S_2}\delta_{T_1 T_2}\delta_{T_{z1}T_{z2}}\nonumber\\
 & \times & Q(l_1 J_1 M_{1},l_2 J_2 M_{2}:S_1 T_1 k K \theta_K \phi_K),
 \end{eqnarray}
where
\begin{eqnarray}
 \label{eq:Q1}
   \lefteqn{
 Q(l_1 J_1 M_{1},l_2 J_2 M_{2}:S T k K \theta_K \phi_K)
            }\hspace{1cm}\nonumber\\
 &=& f_{l_1 S T}\hspace{1mm}f_{l_2 S T}
 \sum_{m_1 m_2 S_{z1},S_{z2}}\int d\hat{\bm{k}} \hspace{1mm}
     Y^*_{l_1 m_1}(\hat{\bm{k}})\hspace{1mm}
     Y_{l_2 m_2}(\hat{\bm{k}})\nonumber\\
 & &\times \< l_1 m_1 S S_{z1}|J_1 M_{1}\>
    \< l_2 m_2 S S_{z2}|J_2 M_{2}\>\nonumber\\
 & &\times \hspace{1mm}\Theta(|\frac{\bm{K}}{2}+\bm{k}|-k_{\rm F})
           \hspace{1mm}\Theta (|\frac{\bm{K}}{2}-\bm{k}|-k_{\rm F}).
\end{eqnarray}
Here $\theta_K$ and $\phi_K$ are the polar angles of {\boldmath $K$}.
Due to the presence of the step function, the solid angle
${\hat {\bm k}}$ is restricted in the integration, and therefore
 $l$ and {\boldmath $J$} are, in general, not conserved  
in the matrix element of the 
 operator $Q$,  while {\boldmath $K$}, $k$, $S$, $T$ and $T_z$  are conserved.

If we employ a reference frame, referred to as the K system 
hereafter, in which {\boldmath $K$} points in the direction of 
the $z$ axis, the angle between 
{\boldmath $K$} and {\boldmath $k$} coincides with the colatitude $\theta$ of
{\boldmath $k$}.
In this case the limit on  the integration with respect to the angle $\theta$
  in Eq.(\ref{eq:Q1}) 
is given by 
$-x_0 \leq {\rm cos}\theta \leq x_0$, where
\begin{equation}
 \label{eq:czero}
   x_0 = \left\{
           \begin{array}{cl}
             0 & \hspace{1mm}\mbox{for}\hspace{5mm}k < \sqrt{k^2_{\rm F}-K^2 / 4},\\
             \frac{K^2 /4 + k^2-k^2_{\rm F}}{K k}
               & \mbox{    for}\hspace{5mm}
                    \sqrt{k^2_{\rm F}-K^2 / 4}< k <k_{\rm F}+K/2,\\
             1 & \hspace{1mm}\mbox{otherwise.}
           \end{array}
          \right.
 \end{equation}
Since  the integration with respect to the longitude $\phi$ 
of the relative momentum {\boldmath $k$} can be carried out in the K system, 
 the matrix element of the exclusion operator, denoted by $Q_0$, is written as
\begin{eqnarray}
 \label{eq:Q2Ksystem}
  \lefteqn{
  Q_{0}(l_1 J_1 M_{1},l_2 J_2 M_{2}:S T kK)
          }\nonumber\\
 &=& f_{l_1 S T} \hspace{1mm} f_{l_2 S T}
   \hspace{1mm}\delta_{M_{1} M_{2}}
   \sum_{L} (-1)^{S+M_{1}}
     \hat{l_1}\hat{l_2}\hat{J_1}\hat{J_2}
   \< l_1 0 l_2 0|L 0 \>
   \< J_1\hspace{1mm}{-M}_{1} J_2 M_{1}|L 0\>\nonumber\\ 
   & &\times W(l_1 J_1 l_2 J_2 ; S L)
       \int^{x_0}_{0} P_L (x)dx,
\end{eqnarray}
where $P_L (x)$ is the Legendre polynomial, and 
we have used the notation $\hat{l}\equiv \sqrt{2 l +1}$.
It is clear from the above expression of the operator $Q$ that, in the K system,
 the projection  $M$  is 
conserved  though the magnitude of {\boldmath $J$} is not.
It might be useful to show that the matrix element $Q_{0}(l_1 J_1 M,l_2 J_2 M:S T kK)$
satisfies the closure relation

\begin{equation}
\label{eq:Qclosure}
 \frac{1}{2J_1 +1}\sum_M Q_{0}(l_1 J_1 M,l_2 J_2 M:S T kK)\delta_{J_1 J_2}=
   f_{l_1 S T} \hspace{1mm} x_0 \delta_{J_1 J_2}\delta_{l_1 l_2}.
\end{equation}

We next consider the matrix element of the operator $Q$ in an arbitrary
 reference frame in which 
the direction of {\boldmath $K$} does not coincide with the $z$ axis.
By making rotation of the reference frame through the Euler angles 
$\aa=\phi_K, \bb=\theta_K$ and $\cc=0$, we obtain a general expression of 
the matrix element of $Q$ as 
\begin{eqnarray}
 \label{eq:Q2gen}
      \lefteqn{
 Q(l_1 J_1 M_{1},l_2 J_2 M_{2}:S T k K \theta_K \phi_K )
              }\hspace{0.5cm}\nonumber\\
  &=& \sum_{M'} D^{J_1}_{M_1 M'}(\theta_K,\phi_K,0)
             {D^{J_2 *}_{M_2 M'}}(\theta_K,\phi_K,0)
             Q_{0}(l_1 J_1 M',l_2 J_2 M':S T kK)
             \nonumber\\
  &=& f_{l_1 S T}\hspace{1mm}f_{l_2 S T}
     \sum_{L} (-1)^{S+M_{1}}
     \frac{\sqrt{4\pi}\hat{l_1}\hat{l_2}\hat{J_1}\hat{J_2}}{\hat{L}}
                          \< l_1 0 l_2 0|L 0 \>
   \< J_1 \hspace{1mm}{-M}_{1} J_2 M_{2}|L M \>\nonumber\\
  & &\times Y_{LM}(\theta_K,\phi_K) W(l_1 J_1 l_2 J_2 ; S L)
   \int_0^{x_0}P_L (x)dx,
\end{eqnarray}
where  the function $D^{J}_{MM'}(\aa,\bb,\cc)$ is the Wigner 
$D$-function\cite{rf:VMK89} of the Euler angles 
$(\aa,\bb,\cc)$. In the derivation of
 Eq.(\ref{eq:Q2gen}) 
 we have used the fact that only even integers $L$ are allowed 
 due to the parity conservation. We remark here
 that the matrix element of the exclusion operator 
  is factorized as 
 \begin{eqnarray}
 \label{eq:qsolution}
  \lefteqn{
  Q(l_1 J_1 M_{1},l_2 J_2 M_{2}:S T k K \theta_K \phi_K )
          }\hspace{10cm}\nonumber\\
  =Q(l_1 J_1 M_{1},l_2 J_2 M_{2}:S T k K \theta_K \phi_K =0)
   \times {\rm e}^{i(M_2 - M_1)\phi_K}.
 \end{eqnarray}
The above expression shows clearly  how the matrix element of the
operator $Q$  depends on the angle $\phi_K$.

By using the recurrence formulae for the Legendre polynomials 
we obtain an integral formula for an even integer $L$ as
\be
 \label{eq:int} 
  \int_0^{x_0}P_{L}(x)dx=\frac{1}{2L+1}[ P_{L+1}(x_0)-P_{L-1}(x_0)].
\ee
With use of the above relation we may rewrite Eq.(\ref{eq:Q2gen}) as
\begin{eqnarray}
 \label{eq:Q2gen2}
\lefteqn{
 Q(l_1 J_1 M_{1},l_2 J_2 M_{2}:S T k K \theta_K \phi_K )}\nonumber\\
  &=& f_{l_1 S T}\hspace{1mm}f_{l_2 S T}
    \{ x_0\delta_{l_1 l_2}\delta_{J_1 J_2}\delta_{M_{1} M_{2}}\nonumber\\
  &   +& \sum_{L>0,L={\rm even}} (-1)^{S+M_{1}}
     \frac{\sqrt{4\pi}\hat{l_1}\hat{l_2}\hat{J_1}\hat{J_2}}
          {\hat{L}^3}
                          \< l_1 0 l_2 0|L 0 \>
   \< J_1 \hspace{1mm}{-M}_{1} J_2 M_{2}|L M \>\nonumber\\
  & &\times Y_{LM}(\theta_K,\phi_K) W(l_1 J_1 l_2 J_2 ; S L)
       [ P_{L+1} (x_0)-P_{L-1}(x_0)] \}.
\end{eqnarray}
It might be useful to give another expression of  Eq.(\ref{eq:Q2gen}) written as
\begin{eqnarray}
 \label{eq:Q2gen3}
\lefteqn{
 Q(l_1 J_1 M_{1},l_2 J_2 M_{2}:S T k K\theta_K \phi_K)
        }\hspace{0.5cm}\nonumber\\
  &=&f_{l_1 S T}\hspace{1mm}f_{l_2 S T}\hspace{1mm}
     [ x_0\delta_{l_1 l_2}\delta_{J_1 J_2}\delta_{M_{1} M_{2}}\nonumber\\
  &   +& \sum_{L>0,L={\rm even}} (-1)^{S+M_{1}}
     \frac{\sqrt{4\pi}\hat{l_1}\hat{l_2}\hat{J_1}\hat{J_2}}
          {\hat{L}}
                          \< l_1 0 l_2 0|L 0 \>
   \< J_1  \hspace{1mm}{-M}_{1} J_2 M_{2}|L M \>\nonumber\\
  & &\times Y_{LM}(\theta_K,\phi_K) W(l_1 J_1 l_2 J_2 ; S L)
       \frac{1}{L(L+1)}(x_0^2 - 1) P'_L (x_0)],
\end{eqnarray}
where $P^{'}_L (x_0)$ is the first derivative of $P_L (x)$ at $x=x_0$.
It is easy to see from the above expression of the 
operator $Q$ that if $x_0 = 1$, the second term on the right-hand side 
of Eq.(\ref{eq:Q2gen3}) vanishes.
 Therefore if $k>k_F +K/2$, in this case $x_0 = 1$ as 
given in Eq.(\ref{eq:czero}),
 the matrix element of the
operator $Q$ conserves $l, J$ and $M$ and becomes unity.
We emphasize that
 Eqs. (\ref{eq:Q2gen2}) and (\ref{eq:Q2gen3})
 do not include integrals or step functions anymore, and  are
 analytic and general expressions of the matrix element of $Q$
 which are valid for any vector {\boldmath $K$}.

We discuss here the angle-average approximation for the operator $Q$,
which was first introduced by  Brueckner and Gammel\cite{rf:BG58}
and has been adopted in many nuclear matter calculations.
The operator $Q$ in the angle-average approximation
 is defined as an average over the angle of {\bm K} as
\begin{eqnarray}
 \label{eq:aadef}
 \lefteqn{
  \overline{Q}(l_1 J_1 M_1 ,l_2 J_2 M_2:S T kK)
          }\hspace{10cm}\nonumber\\
 \hspace{5mm}\equiv\frac{1}{4\pi}\int d{\hat {\bm{K}}}\hspace{2mm}
 Q(l_1 J_1 M_{1},l_2 J_2 M_{2}:S T k K \theta_K \phi_K ).
\end{eqnarray}
In this integration the second term on the right-hand side of
Eq.(\ref{eq:Q2gen2}) or (\ref{eq:Q2gen3}) vanishes, and we have
\be
\overline{Q}(l_1 J_1 M_1,l_2 J_2 M_2:S T kK)
 = {f_{l_1 S T}}\hspace{1mm}
  x_0 \delta_{l_1 l_2}\delta_{J_1 J_2}\delta_{M_1 M_2}.
\ee
It is obvious that the first term on the right-hand side of 
Eq.(\ref{eq:Q2gen2}) or (\ref{eq:Q2gen3}) 
gives the angle-averaged $Q$ operator and the second term provides
 the correction which comes from non-spherical characters of 
the operator $Q$.

\section{One-body potential and ground-state energy}
\setcounter{equation}{0}

The  $G$ matrix equation in the lowest-order Brueckner theory  is given 
in angular-momentum-coupling states by
\begin{eqnarray}
 \label{eq:gmateq}
  \lefteqn{
 \< k_1 l_1 S J_1 M_1 T|G(\omega, K \theta_K \phi_K)
   |k_2 l_2 S J_2 M_2 T\>
          }
 \hspace{2cm}\nonumber\\
  =\<&k_1& l_1 J_1 ST|v|k_2 l_2 J_2 ST \>
      \delta_{J_1 J_2}\delta_{M_1 M_2}\hspace{4cm}\nonumber\\
  &+&\sum_{\stackrel{l'_1 l'_2 J'_1 J'_2}
                      {M'_1 M'_2}}
     \int k^2 dk \< k_1 l_1 J_1 ST|v|k l'_1 J'_1 ST \>
                   \delta_{J_1 J'_1}\delta_{M_1 M'_1}\nonumber\\
  &      \times&Q(l'_1 J'_1 M'_1 l'_2 J'_2 M'_2:ST kK\theta_K \phi_K)
   \times\frac{1}{\omega - T(K,k)}\nonumber\\
  &      \times&\
       \< k    l'_2 S J'_2 M'_2 T|G(\omega, K \theta_K \phi_K)
         |k_2  l_2  S J_2  M_2  T\>,
\end{eqnarray}
where   $\omega$ is the starting energy and $T(K,k)$ the kinetic energy defined as 
\be
  T(K,k)=\frac{K^2}{4m}+\frac{k^2}{m}
\ee
with the nucleon mass $m$. We adopt the conventional $QTQ$ spectra for the 
intermediate energies in the present work.

The $G$ matrix is generally diagonal in the c.m.
momentum {\boldmath $K$}, the spin $S$ and the isospin $T$,
 but not in the angular momentum
{\boldmath $l$} and the coupled angular momentum {\boldmath $J$} of relative states.
Corresponding to the factorization of the matrix element
 of the operator $Q$ 
as shown in Eq.(\ref{eq:qsolution}), the $G$ matrix is factorized as
\begin{eqnarray}
 \label{eq:gphase}
  \lefteqn{
 \< k_1 l_1 S J_1 M_1 T|G(\omega, K \theta_K \phi_K)
   |k_2 l_2 S J_2 M_2 T\> }\hspace{12cm}\nonumber\\
 =\< k_1 l_1 S J_1 M_1 T|G(\omega, K \theta_K \phi_K =0)
   |k_2 l_2 S J_2 M_2 T\>\times {\rm e}^{i(M_2 - M_1)\phi_K}.
 \end{eqnarray}
This relation expresses explicitly how the $G$ matrix depends 
on the angle $\phi_K$.
It is to be noted that the $G$ matrix is not invariant under
the rotation about the $z$ axis, but the rotation yields
a phase factor as given in Eq.(\ref{eq:gphase}).

The ground-state energy per nucleon of nuclear matter is given
by
\begin{eqnarray}
 \label{eq:gse}
  E/A &=&\sum_{\lambda}\< \lambda|t|\lambda\>
       +\frac{1}{2}\sum_{\lambda\mu\leq\rho_{\rm F}}
         \< \lambda\mu |G|\lambda\mu \>\nonumber\\
      &=& \frac{3}{5}(\frac{\hbar^2}{2m}k^2_{\rm F})
         +\frac{3}{2k^3_{\rm F}}
           \int_{0}^{k_{\rm F}} u(k_{\lambda})k^2_{\lambda}dk_{\lambda},
\end{eqnarray}
where $t$ is the one-body kinetic energy, $u(k_{\lambda})$  the
 potential energy of a nucleon in the occupied state $|\lambda\>$,
 and $\rho_{\rm F}$ the upper most occupied single-particle level
 (the Fermi level).
Without loss of generality one can choose a reference frame 
 in which the z axis (the quantization axis)
coincides with the direction of the momentum  {\boldmath $k$}$_{\lambda}$
as {\boldmath $k$}$_{\lambda}=(0, 0, k_{\lambda})$.
We call this frame the L system hereafter. The self-consistent 
potential $u(k_{\lambda})$
 is  written in terms of the $G$ matrix in the L system as
\begin{eqnarray}
 \label{eq:potL}
  u(&k_{\lambda}&)=\sum_{\mu\leq\rho_{\rm F}}
                    \< \lambda\mu|G|\lambda\mu\>\nonumber\\
    &   &=\sum_
      {\stackrel
        {\scriptstyle l_1 J_1 l_2 J_2 ST}
        {\scriptstyle m_1 m_2 S_z M_{1} M_{2}}}
    \frac{(2T+1)}{2}
    \int_0^{k_{\rm F}} k_{\mu}^2 dk_{\mu}\int_{-1}^1 d{\rm cos}\theta_{\mu}
    \int_0^{2\pi}d\phi_{\mu}\nonumber \\
    & &\times  \< k_{\lambda\mu} (l_1 S)J_1 M_{1}T|
      G(\omega, K\theta_K \phi_K)| k_{\lambda\mu} (l_2 S)J_2 M_{2}T\>
       \< l_1 m_1 SS_z|J_1 M_{1}\>\nonumber\\
    & &\times \< l_2 m_2 SS_z|J_2 M_{2}\>
       Y_{l_1 m_1}(\theta_{\lambda\mu},\phi_{\lambda\mu})
       Y_{l_2 m_2}^* (\theta_{\lambda\mu},\phi_{\lambda\mu}),
\end{eqnarray}
where $\omega$ is the starting energy given by
\be
 \label{eq:omega}
 \omega= \<\lambda|t|\lambda\>+u(k_{\lambda})
        +\<\mu|t|\mu\>+u(k_{\mu}).
\ee
Here the angles ($\theta_{\mu}$, $\phi_{\mu}$) and 
($\theta_{\lambda\mu}$, $\phi_{\lambda\mu}$)
are the polar angles of {\boldmath $k$}$_{\mu}$
 and  {\boldmath $k$}$_{\lambda\mu}$  
in the L system, respectively,
where {\boldmath $k$}$_{\lambda\mu}$ is the relative momentum defined by 
{\boldmath $k$}$_{\lambda\mu}=$
({\boldmath $k$}$_{\lambda}-${\boldmath $k$}$_{\mu})/2$. 

The integration in Eq.(\ref{eq:potL}) is much simplified if we 
use the $G$ matrix in the K system in which $\phi_K = \theta_K = 0$.
Let $\< k_{\lambda\mu} (l_1 S)J_1 M T|
      G_0 (\omega, K)| k_{\lambda\mu} (l_2 S)J_2 M T\>$
 be the matrix element of the $G$ matrix in the K system.
Rotating the reference frame through  the Euler angles 
$(\aa=\phi_K,\bb=\theta_K,\cc=0)$, the $G$ matrix in the 
L system is related to that in the K system  as
\vspace*{-0.5cm}
\begin{eqnarray}
 \label{eq:Glk}
 \lefteqn{
 \< k_{\lambda\mu} (l_1 S)J_1 M_{1}T|
      G(\omega, K\theta_k \phi_K)| k_{\lambda\mu}(l_2 S)J_2 M_{2}T\>
         }\hspace{1cm}
   \nonumber\\
 &=&\sum_{M} 
   D^{J_1}_{M_{1}M}(\phi_K,\theta_K,0) 
   D^{J_2 *}_{M_{2}M}(\phi_K,\theta_K,0)\nonumber\\
 &  \times& \< k_{\lambda\mu} (l_1 S)J_1 M T|
      G_0 (\omega, K)| k_{\lambda\mu} (l_2 S)J_2 M T\> .
\end{eqnarray}

Substituting Eq.(\ref{eq:Glk}) into Eq.(\ref{eq:potL}) and making integration
with respect to the angle $\phi_{\mu}$ we have
\vspace*{-0.5cm}
\begin{eqnarray}
 u(k_{\lambda})&=& \sum_{\scriptstyle l_1 J_1 l_2 J_2 MST}
    \frac{2\pi(2T+1)}{2}
         \int_{0}^{k_{\rm F}} k_{\mu}^2dk_{\mu}
         \int_{-1}^{1} d{\rm cos}\theta_{\mu}
      \nonumber\\
    & &\times \< k_{\lambda\mu} (l_1 S)J_1 M T|
      G_{0}(\omega, K)| k_{\lambda\mu} (l_2 S)J_2 M T\>\nonumber\\
    & &\times F(k_{\lambda}k_{\lambda\mu}K(l_1 J_1)(l_2 J_2)STM),
 \label{eq:potfinal}
\end{eqnarray}
where we have  used the facts that  
$\phi_K = \phi_{\mu}$ and $\phi_{\lambda\mu}=\phi_{\mu}+\pi$ in the K system.
The coefficient $F$ in Eq.(\ref{eq:potfinal}) is given by
\begin{eqnarray}
  \lefteqn{
      F(k_{\lambda}k_{\lambda\mu}K(l_1 J_1)(l_2 J_2)STM)
         }\nonumber\\
  &=&\sum_{m_1 m_2 S_z M_1 M_2}(-1)^{m_1 - m_2}
          D^{J_1}_{M_1 M}(0,\theta_K,0)D^{J_2}_{M_2 M}(0,\theta_K,0)
     \nonumber\\
  & &\times \< l_1 m_1 S S_Z  |J_1 M_1 \> \< l_2 m_2 S S_Z  |J_2 M_2 \>
          Y_{l_1 m_1}(\theta_{\lambda\mu},0)
          Y^{*}_{l_2 m_2}(\theta_{\lambda\mu},0).
 \label{eq:Fcoeff1}
\end{eqnarray}
Using some formulae of the Wigner $D$-function, the coefficient $F$
becomes 
\begin{eqnarray}
 \lefteqn{
  F(k_{\lambda}k_{\lambda\mu}K(l_1 J_1)(l_2 J_2)STM)
         }\nonumber\\
  &=&\sum_{L} \frac{1}{4\pi}(-1)^{S+M}\hspace{1mm}
      {\hat l_1}{\hat l_1}{\hat J_1}{\hat J_2}\hspace{1mm}
      W(l_1 J_1 l_2 J_2;SL)\< J_1 \hspace{1mm}{-M} J_2 M | L 0\> \nonumber\\
  & &\times \<l_1 0 l_2 0| L 0\>\hspace{1mm}
      P_L ({\rm cos}(\theta_K+\theta_{\lambda\mu})).
 \label{eq:Fcoeff2}
\end{eqnarray}
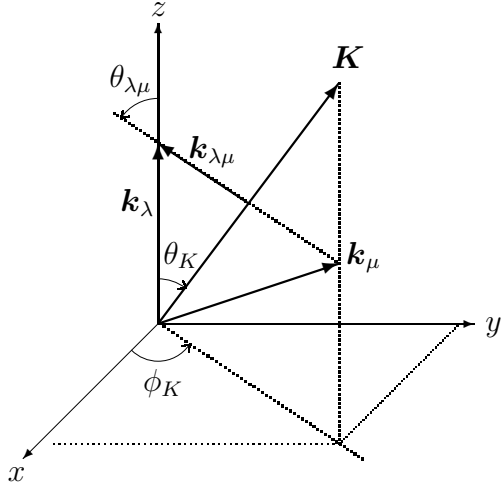
\begin{figure*}
  \label{fig:1}
{
\setlength{\unitlength}{1mm}
\begin{picture}(90,70)(-40,10)
\put(20,30){\vector(1,0){42}}
\put(20,30){\vector(0,1){40}}
\put(20,30){\vector(-1,-1){18}}

\put(0,9){\mbox{\(x\)}}
\put(63.5,29){\mbox{\(y\)}}
\put(19,71){\mbox{\(z\)}}
\put(18,21){\mbox{$\phi_K$}}
\put(20.7,38){\mbox{$\theta_K$}}
\put(13.5,61.5){\mbox{$\theta_{\lambda \mu}$}}
 \put(43,64){\mbox{\boldmath{$K$}}}
\put(15,45){{\mbox{\boldmath{$k$}}$_\lambda$}}
\put(45,38){{\mbox{\boldmath{$k$}}$_\mu$}} 
\put(24,52){{\mbox{\boldmath{$k$}}$_{\lambda \mu}$}} 

\put(20,30){\arc{10}{0.588}{2.356}}
\put(20,30){\arc{12}{4.712}{5.356}}
\put(20,54){\arc{12}{3.730}{4.712}}

\bezier{8}(23.6,34.8)(23.1,34.7)(22.6,34.6) 
\bezier{8}(23.6,34.8)(23.4,35.35)(23.2,35.7) 
\bezier{8}(15.0,57.33)(15.1,57.83)(15.2,58.33)
\bezier{8}(15.0,57.33)(15.5,57.43)(16.0,57.53) 
\bezier{8}(24.16,27.23)(23.66,27.13)(23.16,27.03) 
\bezier{8}(24.16,27.23)(24.06,26.73)(23.96,26.23) 

\bezier{50}(6,14)(25,14)(44,14)
\bezier{50}(44,14)(52,22)(60,30)

\thicklines
\put(20,30){\vector(3,1){24}}
\put(20,30){\vector(0,1){24}}
\put(20,30){\vector(3,4){24}}
\put(32,46){\vector(-3,2){12}}
\bezier{50}(20,30)(35,20)(47,12)
\bezier{60}(44,38)(29,48)(14,58)
\bezier{80}(44,14)(44,38)(44,62)
\end{picture}
}
  \caption {The relation among relevant momentum vectors.}
\end{figure*}
 
As shown in Fig.1,  $\theta_K+\theta_{\lambda\mu}$ is the
 angle between the relative  momentum {\boldmath $k$}$_{\lambda\mu}$ and the
c.m. momentum  {\boldmath $K$}, which satisfies
\begin{equation}
{\rm cos}(\theta_K+\theta_{\lambda\mu})
 =\frac{ k_{\lambda}^2 - k_{\lambda\mu}^2-K^2 /4 }
       {k_{\lambda\mu} K}.
\end{equation}

It is clear from Eq.(\ref{eq:potfinal}) that the single-particle
 potential $u(k_{\lambda})$ is determined by the $G$ matrix 
in the K system and the $F$ coefficient 
which is a function of the angle $\theta_K+\theta_{\lambda\mu}$ between 
{\boldmath $k$} and {\boldmath $K$}.

In the angle-average approximation the $G$ matrix in the K system
is  independent of the quantum number $M$.
Using  the following closure relation of the coefficient $F$
\be
 \sum_{M} F(k_{\lambda}k_{\lambda\mu}K(l_1 J_1)(l_2 J_2)STM)\delta_{J_1 J_2}
 =\frac{2J_1 +1}{4\pi}\delta_{J_1 J_2}\delta_{l_1 l_2},
\ee
the single-particle potential $u(k_{\lambda})$ with the angle-averaged $Q$ 
operator becomes
\begin{eqnarray}
 \label{eq:tabakin}
 u(k_{\lambda})&=& \sum_{\scriptstyle l J ST}
    \frac{(2T+1)(2J +1)}{4}
     \int_{0}^{k_{\rm F}} k_{\mu}^2dk_{\mu}
         \int_{-1}^{1} d{\rm cos}\theta_{\mu}\nonumber\\
    & &\times \< k_{\lambda\mu} (l S)J T|
      G_{0}(\omega, K)| k_{\lambda\mu} (l S)J T\>.
\end{eqnarray}
This expression of $u(k_{\lambda})$ agrees with the usual formula
 in the angle-average approximation for the operator $Q$.

\section{Numerical calculation}
\setcounter{equation}{0}

In order to examine the effect of the exact treatment of 
the Pauli exclusion operator  we performed
a numerical calculation of the ground-state properties of nuclear matter
by adopting the Bonn B and C
{\it NN} potentials\cite{rf:Bonn}.
We solved self-consistently the coupled equation (\ref{eq:gmateq})
 for the $G$ matrix in the
K system where the c.m. momentum {\bm K} points in the $z$ direction.
 In the calculation we took
into consideration rigorously the contributions of the partial waves of 
$J\leq 6$. Other higher partial waves 
were taken up to $ J = 18 $ in the Born approximation.
We checked the stability of the calculated result with respect
 to the number of  mesh points in numerical integration.
Furthermore we confirmed that 
the same result was obtained within  numerical errors by using
two  computer codes made independently.  

The calculated results of the binding energies per nucleon
are presented in Table~1 and shown in Figs.~2 and 3, respectively.
From Table~1 it is  seen, as a  common characteristic of the results for 
 two {\it NN} potentials,  
that the exact treatment of the operator $Q$ brings about
 attractive contributions
 to the binding energy per nucleon at any nuclear densities,
compared with the result in the standard angle-average approximation.
As is shown in Figs.~2 and 3 the saturation densities hardly change in the exact
treatment of the operator $Q$. 

\pagebreak
\noindent
{\footnotesize
Table~1. Calculated results of the binding energy per nucleon 
 with the exact and angle-averaged $Q$ operators for the 
 Bonn B and C {\it NN} potentials\cite{rf:Bonn}. 
The energies are in  MeV.
}
\\
\\
\\
{\footnotesize (1) Bonn B {\it NN} potential.}
{\small
  \begin{center}
    \begin{tabular}{crrrrrrr}
                               \hline
$k_{\rm F}[{\rm fm}^{-1}]$& 1.2   &  1.3  & 1.4 &  1.5   & 1.6  & 1.7 & 1.8 \\
                              \hline
exact      &$-$10.28&$-$11.72&$-$12.98&$-$13.95&$-$14.47&$-$14.42&$-$13.61\\
average    &$-$10.18&$-$11.59&$-$12.82&$-$13.75&$-$14.25&$-$14.16&$-$13.32\\
difference & $-$0.10& $-$0.13& $-$0.16& $-$0.20& $-$0.22& $-$0.26& $-$0.29\\
                               \hline
 \end{tabular}
  \end{center}
}
{\footnotesize (2) Bonn C {\it NN} potential.}
{\small
  \begin{center}
    \begin{tabular}{crrrrrrr}
                               \hline
$k_{\rm F}[{\rm fm}^{-1}]$& 1.2  &  1.3  & 1.4 &  1.5   & 1.6  & 1.7 & 1.8 \\
                              \hline
exact      &$-$9.62&$-$10.82&$-$11.78&$-$12.37&$-$12.46&$-$11.90&$-$10.53\\
average    &$-$9.52&$-$10.69&$-$11.62&$-$12.18&$-$12.24&$-$11.65&$-$10.24\\
difference &$-$0.10& $-$0.13& $-$0.16& $-$0.19& $-$0.22& $-$0.25& $-$0.29\\
                               \hline
 \vspace{4mm}
 \end{tabular}
  \end{center}
}
\vspace{10mm}
\begin{figure*}[b]
  \begin{minipage}{2.40in}
    \epsfxsize=6cm
    \centerline{\epsfbox{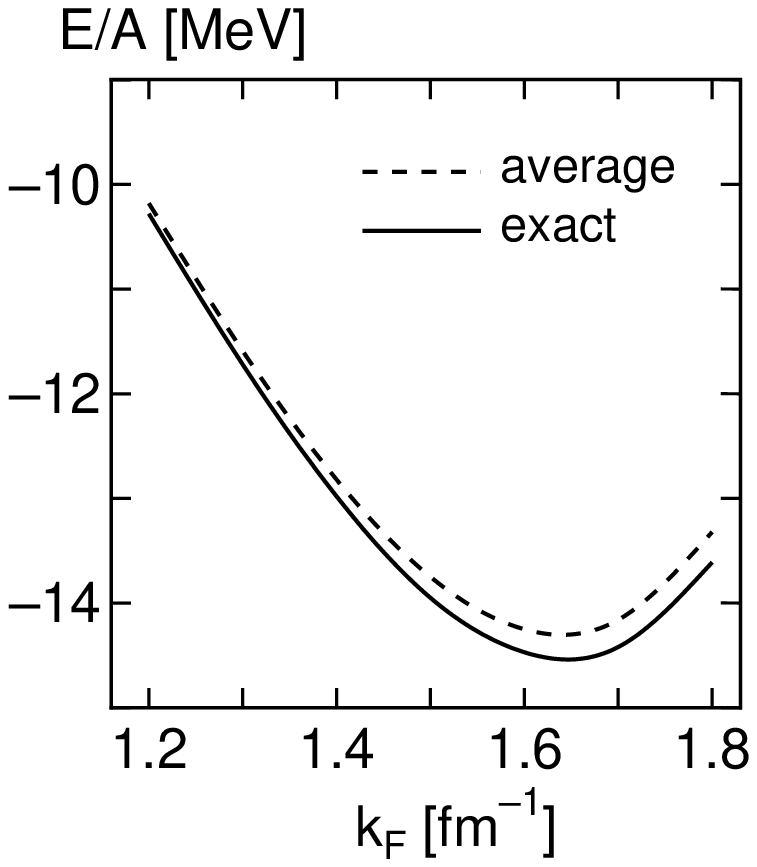}}
    \caption{Calculated binding energies per nucleon as a function of the Fermi momentum
 $k_{\rm F}$ for the Bonn B potential.} 
     \label{fig:bebonnb}
  \end{minipage}
    \hfill
  \begin{minipage}{2.40in}
       \epsfxsize=6cm
    \centerline{\epsfbox{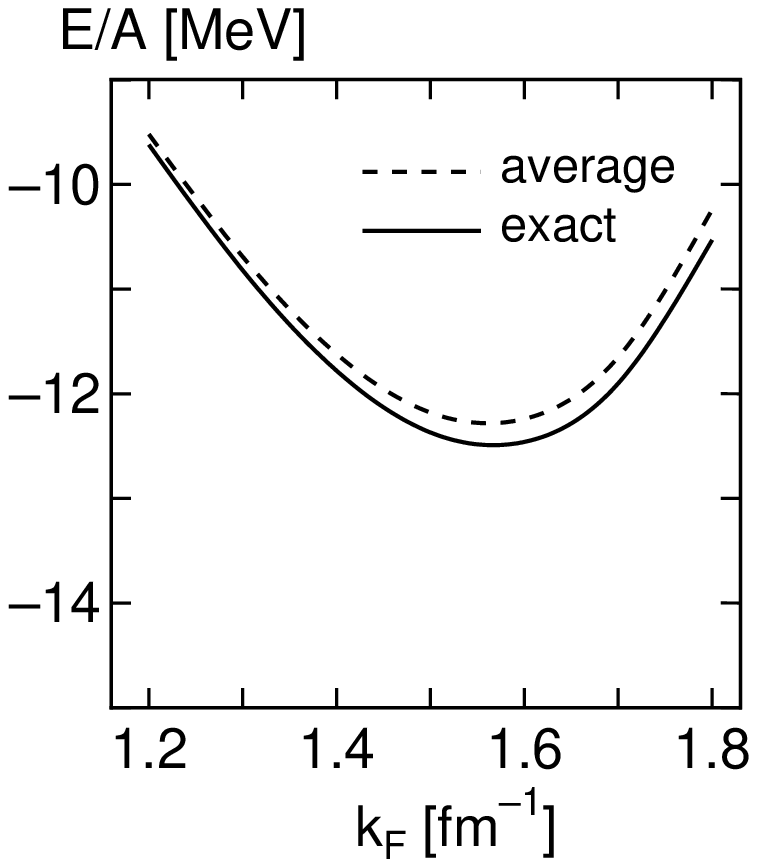}}
    \caption{Calculated binding energies per nucleon as a function of the Fermi momentum
 $k_{\rm F}$ for the Bonn C potential.} 
     \label{fig:bebonnc}
  \end{minipage}
\end{figure*}
 These observations in the present work are 
 quite different from those obtained in a very recent study  by Schiller,
M\"uther and Czerski\cite{rf:SMC99}. 
Their results show that the exact treatment of the operator $Q$ brings about
  repulsive contributions at higher densities, while small attractive
contributions were obtained at low densities. This characteristic density
dependence of the effect from the exact treatment of the operator $Q$ leads
to a non-negligible shift in the calculated saturation point.

In order to analyse  the 
features of our calculated results in more detail, 
we introduce an  approximate decomposition of partial waves
in the exact treatment of the operator $Q$.
Since the $G$ matrix
 in Eq.(\ref{eq:potfinal}) in the K system, $G_0 (\omega, K)$, is not
diagonal in $l$ and $J$, as was discussed in the preceding section,
 the partial wave decomposition is not possible 
in a usual sense in the exact treatment of the operator $Q$. 
However, even in this case, 
we may decompose the contributions to the total potential energy per nucleon
 into two groups of terms which  are 
diagonal and  non-diagonal in $l$ and $J$.
We expect that the partial wave decomposition can be physically meaningful when
the magnitudes of the contributions of the $lJ$-nondiagonal terms are 
sufficiently small
in comparison with the ones of the $lJ$-diagonal terms. We call the diagonal term
in $l$ and $J$ the partial wave contribution of the channel
$(S, T, l, J)$. The summation over $M$ is taken in each of the partial
 wave contributions.

 We show in Table~2 the partial wave contributions of the channels with
lower $l$ and $J$, the sum of the $lJ$-nondiagonal terms and 
the higher partial wave contributions which are treated in the Born 
approximation. The contributions of the $lJ$-nondiagonal
 terms are  shown to be
indeed small in the numerical calculation.
The net contributions of the terms non-diagonal in $l$ and $J$ are
estimated  to be at most 0.4\% of the total potential energy.
 Therefore we may say that the
partial wave analysis could be meaningful even in the
exact treatment of the operator $Q$.

We observed from Table~2 
that about sixty percent of the difference between the exact
 and approximate treatments of the operator $Q$ 
comes from the $lJ$-nondiagonal terms and the 
remainder does from the $lJ$-diagonal terms. We also note that the $P$ waves
 give the most important and attractive contributions.
The  $S$ waves give
the secondary important and repulsive contributions to the difference.
The difference becomes, in general, smaller as $l$ (or $J$) increases.
The reason is that  the Born approximation works well in higher partial waves
and the $G$ matrix 
approaches  to the bare {\it NN} potential, which is independent of the operator $Q$.

\clearpage
\noindent
{\footnotesize
Table~2. Partial wave contributions to the binding energy per nucleon 
 with the exact and angle-averaged $Q$ operators for the 
Bonn B {\it NN} potential\cite{rf:Bonn}. 
The row of ``non-diag" means the sum 
 of the $lJ$-nondiagonal terms. The row of ``higher" is the sum of
 contributions of higher partial waves of $7\leq J \leq 18$ 
treated in the Born approximation. The energies are in  MeV. 
}
{\small
  \begin{center}
\def\arraystretch{1.2}
    \begin{tabular}{crrrrrr}
                                      \hline
$k_{\rm F}[{\rm fm}^{-1}]$  &    1.2 &  1.2   &  1.5 &  1.5  &  1.8  & 1.8  \\  
channel  & exact& average&       exact& average &  exact  & average\\
                                      \hline
$^1 S_0$ & $-$13.018& $-$13.019&$-$20.112& $-$20.117&  $-$26.749& $-$26.760\\
$^1 D_2$ &  $-$1.419&  $-$1.419& $-$3.673&  $-$3.671&  $ -$7.741& $ -$7.737\\
$^1 G_4$ &  $-$0.227&  $-$0.227& $-$0.708&  $-$0.708&  $ -$1.674&  $-$1.674\\
$^1 I_6$ &  $-$0.049&  $-$0.049& $-$0.195&  $-$0.195&  $ -$0.540& $ -$0.540\\
$^3 S_1$ & $-$15.638& $-$15.637&$-$22.195& $-$22.210&  $-$28.058& $-$28.109\\
$^3 D_1$ &     0.909&     0.910&    2.300&     2.301&      4.605&     4.607\\
$^3 D_2$ &  $-$2.393&  $-$2.393& $-$5.933&  $-$5.930&  $-$11.813& $-$11.808\\
$^3 D_3$ &     0.145&     0.147&    0.426&     0.430&      0.920&    0.932\\
$^3 G_3$ &     0.105&     0.105&    0.365&     0.365&      0.931&    0.931\\
$^3 G_4$ &  $-$0.369&  $-$0.369& $-$1.212&  $-$1.212&  $ -$2.957& $ -$2.957\\
$^3 G_5$ &     0.038&     0.038&    0.150&     0.150&      0.418&     0.419\\
$^3 I_5$ &     0.016&     0.016&    0.072&     0.072&      0.222&     0.222\\
$^3 I_6$ &  $-$0.071&  $-$0.071& $-$0.301&  $-$0.300&   $-$0.873&  $-$0.873\\
$^1 P_1$ &     2.982&     2.989&    6.443&     6.462&     11.944&    11.987\\
$^1 F_3$ &     0.492&     0.493&    1.320&     1.320&      2.783&     2.784\\
$^1 H_5$ &     0.104&     0.104&    0.363&     0.363&      0.911&     0.911\\
$^3 P_0$ &  $-$2.552&  $-$2.552& $-$4.535&  $-$4.536&  $ -$6.216& $ -$6.218\\
$^3 P_1$ &     6.690&     6.693&   14.980&    14.988&     28.680&    28.702\\
$^3 P_2$ &  $-$4.423&  $-$4.415&$-$10.859& $-$10.839&  $-$21.637& $-$21.598\\
$^3 F_2$ &  $-$0.305&  $-$0.305& $-$0.908&  $-$0.908&  $ -$2.015& $ -$2.014\\
$^3 F_3$ &     0.876&     0.876&    2.486&     2.486&      5.458&     5.458\\
$^3 F_4$ &  $-$0.148&  $-$0.148& $-$0.579&  $-$0.578&  $ -$1.661& $ -$1.660\\
$^3 H_4$ &  $-$0.040&  $-$0.040& $-$0.160&  $-$0.160&  $ -$0.457& $ -$0.457\\
$^3 H_5$ &     0.159&     0.159&    0.590&     0.590&      1.564&     1.564\\
$^3 H_6$ &  $-$0.016&  $-$0.016& $-$0.077&  $-$0.077&  $ -$0.250& $ -$0.250\\
$^3 J_6$ &  $-$0.007&  $-$0.007& $-$0.034&  $-$0.034&  $ -$0.115&  $ -$0.115\\
higher   &     0.044&    0.044 &  0.204  &     0.204&      0.624  &  0.624\\
non-diag &  $-$0.083&    0.0   & $-$0.156&     0.0  &   $-$0.226&     0.0  \\
Total    & $-$28.197& $-$28.094&$-$41.940& $-$41.747&  $-$53.921& $-$53.630\\
                               \hline
\vspace{4mm}
    \end{tabular}                
  \end{center}
}
\vspace{5mm}
\noindent
{\footnotesize
Table~3. Single-particle potential energies for various $k_{\lambda}$ values 
 at $k_{\rm F} = 1.40[{\rm fm}^{-1}]$ in the  exact and angle-average 
treatments of the operator $Q$ for the 
Bonn B {\it NN} potential\cite{rf:Bonn}.
The other notations are the same as in Table~1.
}
{\small
  \begin{center}
    \begin{tabular}{crrr} 
                            \hline
                             &  $u(k_{\lambda})$ &     &          \\
$k_{\lambda}[{\rm fm}^{-1}]$ & exact  & average & difference \\
                              \hline
                    0.047    &$-$87.96 &$-$88.54  &   0.58 \\
                    0.237    &$-$87.33 &$-$87.84  &   0.51  \\
                    0.533    &$-$84.69 &$-$84.90  &   0.21  \\
                    0.867    &$-$79.33 &$-$79.12  &$-$0.21  \\
                    1.163    &$-$72.69 &$-$72.22  &$-$0.47  \\
                    1.353    &$-$67.71 &$-$67.22  &$-$0.49  \\
                               \hline
\vspace{4mm}
    \end{tabular}
  \end{center}
}
\begin{figure*}[b]
  \label{fig:udif}
  \epsfxsize = 6cm
  \centerline{\epsfbox{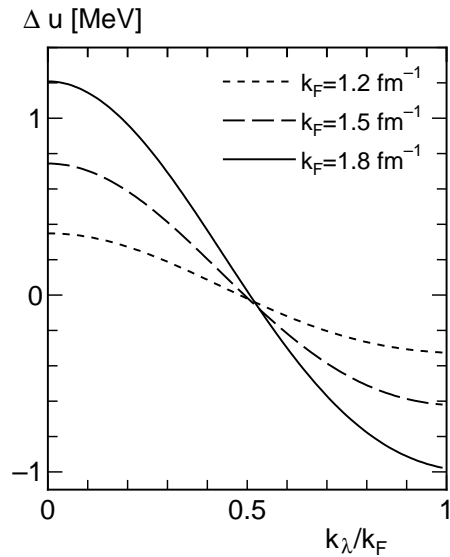}}
  \caption {Differences of $u(k_{\lambda})$'s in the exact and angle-average
 treatments of the operator $Q$ as a function of $k_{\lambda} / k_{\rm F}$
at various nuclear matter densities for the Bonn B potential.
The quantity $\Delta u$ is defined as 
the difference of $u(k_{\lambda})$ with the exact $Q$ from that
 with the angle-averaged $Q$.}
\end{figure*}


Next, we show in Table~3 the single-particle potentials in 
the exact and angle-average treatments of the operator  $Q$ and their differences
for the values of $k_{\lambda}$ of the mesh points in the Gaussian integration. 
The differences  of the single-particle-potential energies 
in the exact treatment from those in the angle-average approximation
are found to be rather small for all the $k_{\lambda}$ values. 
It is remarkable to see that the angle-average 
approximation leads to overestimation of the single-particle-potential energy
 for smaller values of $k_{\lambda}$ and does to underestimation  for 
larger values of $k_{\lambda}$. This feature is observed  
at any nuclear matter densities both for the Bonn B and C potentials.
We display in Fig.~4  the $k_{\lambda}$ dependence of the differences
between two $u(k_{\lambda})$'s in the exact and angle-average treatments of the 
operator $Q$ at three nuclear matter densities.
As shown in Table~3 and Fig.~4 the differences change the sign at about
 $k_{\lambda}=k_F /2$. 
This fact implies that the angle-average approximation gives the correct result
at around $k_{\lambda}=k_F /2$.
Considering  larger phase volumes for larger values of $k_{\lambda}$, 
it is understandable
that the angle-average approximation tends to underestimate
the binding energy per nucleon.

We may conclude from the present study that the 
non-spherical character of the  operator
$Q$  causes a non-negligible and attractive effect on  the nuclear matter
binding energy per nucleon, which has been disregarded in the standard
 angle-average approximation for $Q$. 
We believe that the analytic expression given explicitly in this paper
 for the operator $Q$ in the
$G$ matrix is useful in making a precise numerical calculation
of nuclear matter. 

\section{Concluding remarks}
\setcounter{equation}{0}

We derived an exact and analytic expression of the matrix element of 
the Pauli exclusion operator in nuclear matter, 
and gave its  relation to that in the angle-average approximation.
Furthermore we discussed the breaking  of relevant
 symmetries in the matrix element of the exclusion operator.
We presented the rigorous expressions of the 
single-particle-potential energies and the ground-state energy in a form 
that would be suitable for a precise numerical calculation
in the exact treatment of  the  operator $Q$. 

In order to examine the effect of the exact treatment of the operator $Q$
 and  assess the reliability of the angle-average approximation 
for the operator $Q$ we performed
 numerical calculations of the ground-state properties
of nuclear matter  for various  $k_{\rm F}$ values 
by employing the Bonn B and C {\it NN} potentials.
We found that the exact treatment of the operator $Q$ brought about 
non-negligible and attractive contributions to the nuclear matter binding energy
per nucleon.
Our calculations also clarified the degree to which the angle-average approximation
 is reliable quantitatively. 

\section*{Acknowledgements}
The authors would like to thank T.T.S.~Kuo
for his interest in this study and encouragement.

\end{document}